# Event Detection in Twitter by Weighting Tweet's Features

Parinaz Rahimizadeh, Mohammad Javad Shayegan

Department of Computer Engineering, University of Science and Culture, Tehran, Iran.

parinazrahimizadeh@yahoo.com , shayegan@usc.ac.ir

## ABSTRACT

In recent years, people spend a lot of time on social networks. They use social networks as a place to comment on personal or public events. Thus, a large amount of information is generated and shared daily in these networks. Using such a massive amount of information can help authorities to react to events accurately and timely. In this study, the social network investigated is Twitter. The main idea of this research is to differentiate among tweets based on some of their features. This study aimed at investigating the performance of event detection by weighting three attributes of tweets; including the followers count, the retweets count, and the user location. The results show that the average execution time and the precision of event detection in the presented method improved 27% and 31%, respectively, than the base method. Another result of this research is the ability to detect all events (including hot events and less important ones) in the presented method.

## 1  INTRODUCTION

Social networks are one of the largest sources of information. By the expanse of these networks, the need for systems that are able to extract useful information from such a large amount of data is felt. Twitter is one of these social networks that has received much attention recently. Twitter users can receive and post text messages of up to 280 characters, named tweets. Twitter has 330 million active users monthly who generate more than 500 million tweets daily. Thus, a large amount of information is exchanged in this social network that can be utilized as an essential source for reporting real-world events (Petroviíc et al. 2010; Sakaki et al. 2013; Zhou and Chen 2014; Katragadda et al. 2016).

One of the important characteristics of Twitter is its real-time nature. Twitter users share and discuss different kinds of information, from daily personal events to important and global events, in real-time. Recent studies have shown that reporting and discussing events that users are experiencing them are one of the common usages of social networks. These events may contain critical contents that describe the situations throughout a crisis. Monitoring the critical events, crisis management, and decision making can be done via social streams. These capabilities enable authorities to analyze the general situation of an event and make the right decision (Sakaki et al. 2013), (Dou et al. 2012; Sakaki et al. 2013; Zhou and Chen 2014; Guille and Favre 2015; Avvenuti et al. 2018).

In the event detection domain, we believe that tweets differ from each other in their weights, and all tweets should not be weighed the same. Thus, the main idea of this study is using a hybrid method for weighting tweet's features, including the followers count, the retweets count, and user location.

The followers count: When a user posts a tweet with high followers, it is of higher value than one with fewer followers, and it can be allocated higher weight. Influencer users usually publish accurate information; it means the validity of their posts is higher. These users have more audiences, too. Hence, the news would be distributed more and sooner by them.

The retweets count: When a tweet is retweeted frequently, it means the tweet contains important material that can perhaps be an event-related. Thus, such a tweet has a higher value than a tweet that is not retweeted and can be allocated proportionate weight.

The user location: When the probable location of the tweet's author is near to the probable location of an event, the tweet's value can be considered higher. Locals usually have quick access to the location of the event, and they are able to publish the news and its details sooner. Accordingly, such a tweet can be allocated proportionate weight.

This study aimed to improve event detection performance by weighting the three features mentioned above and utilizing the base method presented in (Kumar et al. 2015).

## 2 LITERATURE REVIEW

Many studies worked on event detection; this section briefly presents the most relevant ones.

Cui et al. (2012) have utilized tags on Twitter as an indicator of events. They have presented a classified algorithm according to three features of hashtags, including instability, Twitter meme possibility, and authorship entropy. Based on these features, hashtags are categorized, and breaking events are detected.

Kaleel and Abhari (2015) have presented a novel method for detecting events from tweet clusters based on locality sensitive hashing. In this method, events detected by matching the event's keywords on cluster labels.

Li et al. (2012) have proposed a scalable segment-based event detection system, named Twevent. This system consists of three main components: I) tweet segmentation, II) event segment detection, and III) event segment clustering.

Katragadda et al. (2016) have presented a new real-time model in which time-evolving graphs were used to detect events in Twitter streams. They also used a topic evolution model for finding credible events and removing noise.

Ozdikis et al. (2012) have introduced a method Based on lexico-semantic expansion of tweets for improving event detection performance on Twitter. The implemented semantic expansion method is based on first-order and second-order (syntagmatic and paradigmatic, respectively) relationships among words.

Boettcher and Lee (2012) have presented a method for detecting local events, called EventRadar. In this method, the average tweet frequency of keywords is estimated per day in and near a potential event zone. These estimations are then used to categorize whether the keywords are local event-related.

Pradhan et al. (2019) have detected events by Bag of Words technique. In this method, a three-phase incremental clustering algorithm was presented for grouping similar tweets effectively. They also offered a heuristics method, named EAAS (Event And Aspects Selection), for detecting an event and its aspects.

McCreadie et al. (2013) have proposed a new scalable method for detecting events. In this method, a new strategy of lexical key partitioning is used to distribute the event detection process among several machines.

Barros et al. (2018) have presented a novel real-time method for detecting events on Twitter, which is based on the entropy calculation of the content of tweets. They used the phase transition of bigrams entropy detection for identifying events.

Phuvipadawat and Murata (2010) have presented a real-time method for detecting and tracking breaking news on Twitter. In this method, researchers have improved grouping results by boosting proper nouns' score. Groups also have been ranked based on popularity, reliability, and freshness factors.

Asadi et al. (2018) have improved the performance of event detection in Twitter streams by considering retweet feature in a thesis that has been done at the University of Science and Culture. The main idea of this study was based on differentiating between tweets and retweets.

Kumar et al. (2015) have introduced a novel method for solving challenges of event detection in real-time Twitter streams. In this study, compression distance and single-pass clustering were used to detect events effectively.

Unankard et al. (2015) have presented a method for early detection of emerging events in social networks, called LSED. In this method, we utilized the mentioned locations in the tweet text for identifying the event's location.

Choi and Park (2019) have presented a method for detecting emerging topics by using High Utility Pattern Mining (HUPM) in Twitter streams. In the HUPM method, both factors of words frequency and word utility are taken into account to detect topics in the pattern generation process.

Nguyen et al. (2019) have introduced a novel method for detecting hot topics on the Twitter data stream. They detected hot topics by incremental clustering, which used named entities and central centroids.

Despite the different studies on event detection, the idea of the hybrid weighting of tweet's features has not been used yet. Hence, this research aimed at investigating event detection performance by the presented idea.

## 3 METHODOLOGY

In this section, research steps, including collecting data, preprocessing data, and event detection process, are presented in detail.

### 3.1 Data Collection

The first stage of the study is collecting data. This study has been done on Twitter. To collecting Twitter data, we used Twitter API that Twitter Company provided. In this study, Persian tweets were investigated, and all data were extracted randomly. At first, about 150000 Persian tweets were studied randomly, and 50 top-most frequent words in these tweets were extracted. After that, the main data were extracted by these 50 words to cover all topics and not just focus on special ones. In this research, Twitter data for six days, including 2019.7.31, 2019.8.1, 2019.8.11, 2019.11.15, 2020.1.3, and 2020.1.8, were extracted. Overall, the Obtained dataset consisted of 600000 tweets.

### 3.2 Data preprocessing

In this research, before entering the primary process of event detection, there is a preprocessing stage that its steps have been shown in Fig. 1.

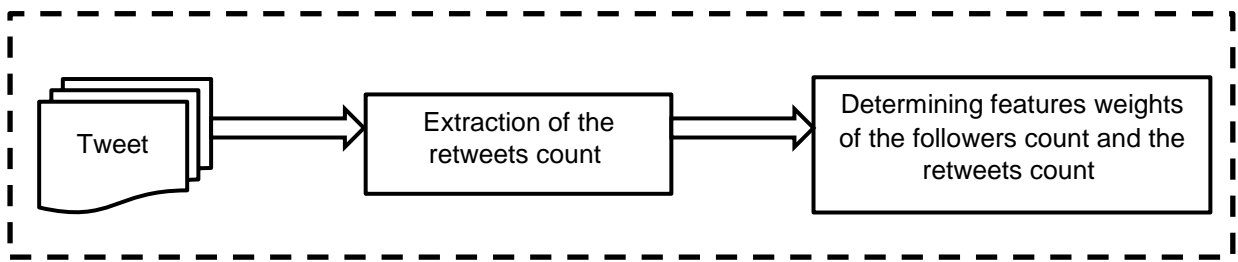

**Fig. 1** Preprocessing of the suggested method

### 3.2.1 Extraction of the retweets count

Data that have been extracted in real-time should be updated by REST API for obtaining the retweets count.

## 3.2.2 Determining features weights of the followers count and the retweets count

For determining the weights of these two features, a sample of tweets about 250000 was first labeled based on whether they are events. After labeling the sample tweets, we utilized the Information Gain method for weighting tweet's features (Zhang et al. 2012).

## 3.3 The primary process of event detection

The suggested event detection method in this study is based on the method in (Kumar et al. 2015). In the primary process of the suggested event detection method, tweets were filtered before clustering. First, URLs were removed from tweets text; then, the tweets contained mentions were ignored and were not clustered. Fig. 2 shows the general framework of the suggested event detection process.

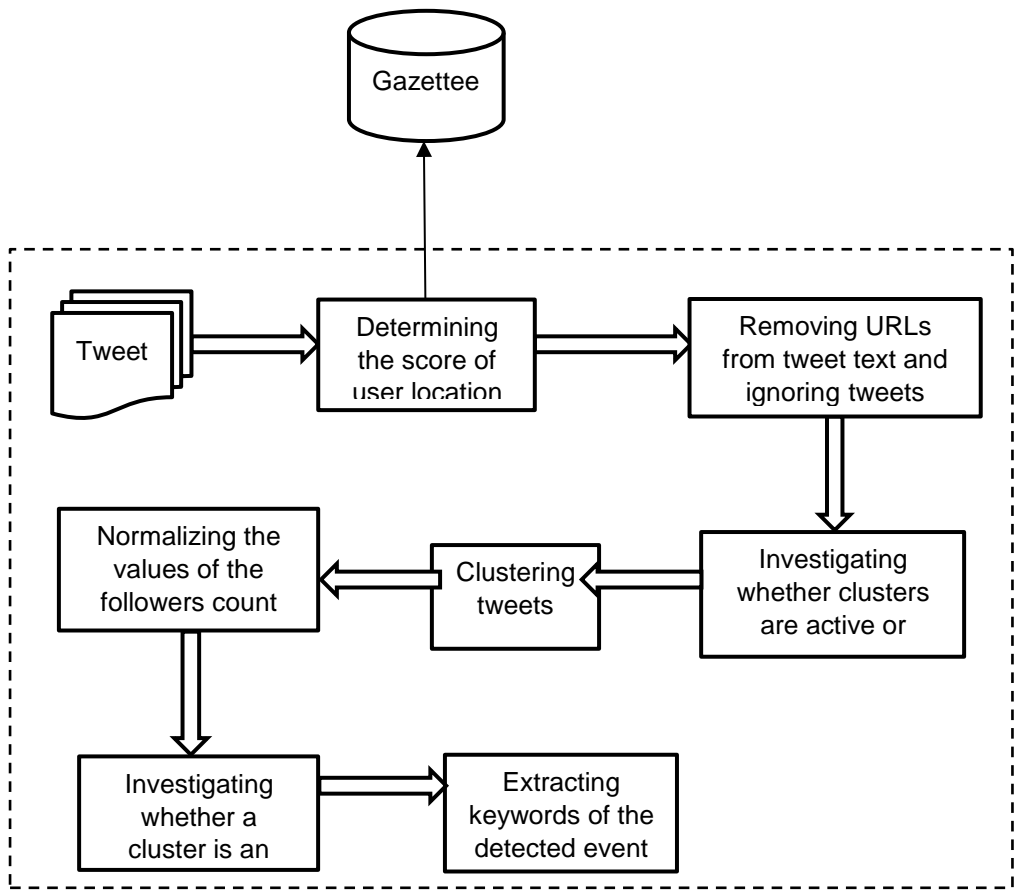

**Fig. 2** General framework of event detection in the suggested method

### 3.3.1 Determining the score of the user location feature

In the suggested method, the score of the user location was obtained by the method used in (Unankard et al. 2015), along with some changes that will be explained in detail.

**Extraction of user location:** In this study, we utilized the registered location in the profile information as the user location.

**Extraction of mentioned locations in the tweet text:** Extracting location from a text is considered as one of the challenging stages in this study. As a result, two methods were combined to get all the mentioned locations in the tweet text. The first method is Named Entity Recognition (NER). The second method is using hashtags in the text. We applied the second method because sometimes the location name is not mentioned in the text of an event-related tweet that is related to a place, whereas it is mentioned in one of the hashtags. Usually, if an event is related to a location, users are more likely to mention that location name in one of the hashtags. Hence, in the current study, a combination of these two methods was applied to extract the mentioned locations in the tweet text.

**User location scoring criterion:** After extracting the user location and the mentioned locations in the tweet text, they were searched in the Gazetteer database, which contains all geographical locations around the world. Finally, the correlation of locations was assigned by the following formula.

$$\text{LocCorrelateScore} = \alpha_1(F(\text{uContinent},\text{eContinent})) + \alpha_2(F(\text{uCountry},\text{eCountry})) \\ + \alpha_3(F(\text{uState},\text{eState})) + \alpha_4(F(\text{uCity},\text{eCity})) \quad (1)$$

In the above formula, $\alpha_1 = \alpha_2 = \alpha_3 = \alpha_4 = 0.25$, and also ucontinent, ucountry, ustate, and ucity are related to the user location and econtinent, ecountry, estate, and ecity are related to the mentioned locations in the tweet text. Comparing the level of the location granularity was used to calculate the correlation score. In this study, granularity level is defined as continent > country > state > city. For calculating this formula, if each level has the same value, it is assigned the score. In other words, if $x=y$, $f(x,y)=1$, otherwise $f(x,y)=0$.

### 3.3.2 Removing URLs from tweet text and ignoring tweets containing mentions

Some tweets have a link in their text that these links have the same format. For this reason, in the base method (Kumar et al. 2015), tweets containing URLs are sometimes mistakenly classified in the same cluster. Therefore, for avoiding such mistakes and improving the cluster's quality, tweets ULRs were removed in this study. Also, studying the data of three days showed that, on average, from the total tweets of a day, only 0.2% of tweets containing mentions were event-related, and most of them contained noisy materials (Fig. 3). Hence, in the current study, for removing such noisy tweets and increasing the speed of tweets processing, tweets containing mentions have not been processed.

### 3.3.3 Investigating whether clusters are active or inactive

Events are dynamic, and considering the temporal evolution of the events on streaming data is necessary. A cluster represents an event that can be considered active or inactive based on the arrival time of a new tweet. In the current study, as in the base method (Kumar et al. 2015), being active or inactive of clusters at any given time was investigated by the passion process. This process is traditionally utilized for modeling the number of objects that are in an event at time *t*. Hence, according to its nature, it can be utilized for estimating the maximum likelihood arrival time of a new tweet to the clusters.

### 3.3.4 Clustering tweets

In this study, when a new tweet enters the system, the value of tweet's distance from other tweets in a cluster can be calculated by compression distance (Keogh et al. 2004). If the calculated value is less than the threshold value ($D_t$) that has been obtained by trial and error, the tweet will be added to that cluster; otherwise, a new cluster will be created and the tweet will be added to it.

In this study, each tweet is considered as a document. *C* is any compressor, *C(x)* is the compressed size of the tweet *x*. Thus, the distance between the two tweets x and y is *D(x,y)* that in the following formula, it has been defined (Kumar et al. 2014):

$$D(x,y) = \frac{C(xy)}{C(x)+C(y)} \quad (2)$$

In the above formula, *C(xy)* is the obtained compression by merging the two tweets.

### 3.3.5 Normalizing the values of the followers count and the retweets count

The followers count and the retweets count values should be normal to be used in the presented formula 3, for investigating whether a cluster is an event. Thus, by the min-max method, the values of these two features were normalized.

### 3.3.6 Investigating whether a cluster is an event

All the detected clusters by the algorithm cannot be events. Thus, it needs to utilize a method for identifying the clusters of events from the other clusters. In the current study, we weighed three features of the tweet for this purpose. The following formula has been presented based on research (Phuvipadawat and Murata 2010).

$$Score_{(c)} = \sum_{i}((W_{followers-count} \times n_{followers-count_i}) + (W_{retweets-count} \times n_{retweets-count_i}) + LocCorrelateScore_i) \quad (3)$$

In the above formula, $n_{followers-count_i}$ is the followers count of the $i^{th}$ tweet's publisher in the cluster, and $n_{retweets-count_i}$ is the retweets count of the $i^{th}$ tweet in the cluster. $W_{followers-count}$, $W_{retweets-count}$ and $LocCorrelateScore_i$ are the weights of the followers count, the retweets count, and the user location features, respectively, that have been obtained in the previous stages. Finally, Formula 3 has been utilized to investigate whether a cluster represents an event or not. When the score of each cluster is higher than the threshold value ($score_t$), that cluster will be considered as an event. The threshold value has been obtained by trial and error.

### 3.3.7 Extracting keywords of the detected event

The discovered events are usually described by the top-most frequent words of its tweets. Hence, in this study, the top keywords of each event were extracted as its description. After that, these words were matched the ground truth to verify the detected event.

## 4 FINDINGS

In this section, the obtained results from testing the suggested method are analyzed.

### 4.1 Results of studying the tweets containing mentions

Based on Fig. 3 and studying the tweets containing mentions during three days of the dataset, the tweets containing mentions comprised only 3.4% of total tweets of a day on average. Also, the average number of tweets containing mentions that were event-related comprised only 0.2% of total tweets of a day. Thus, it can be concluded that removing the tweets containing mentions will have no great negative impact on event detection. However, removing them helps noisy data to be removed significantly, which makes the event detection process faster.

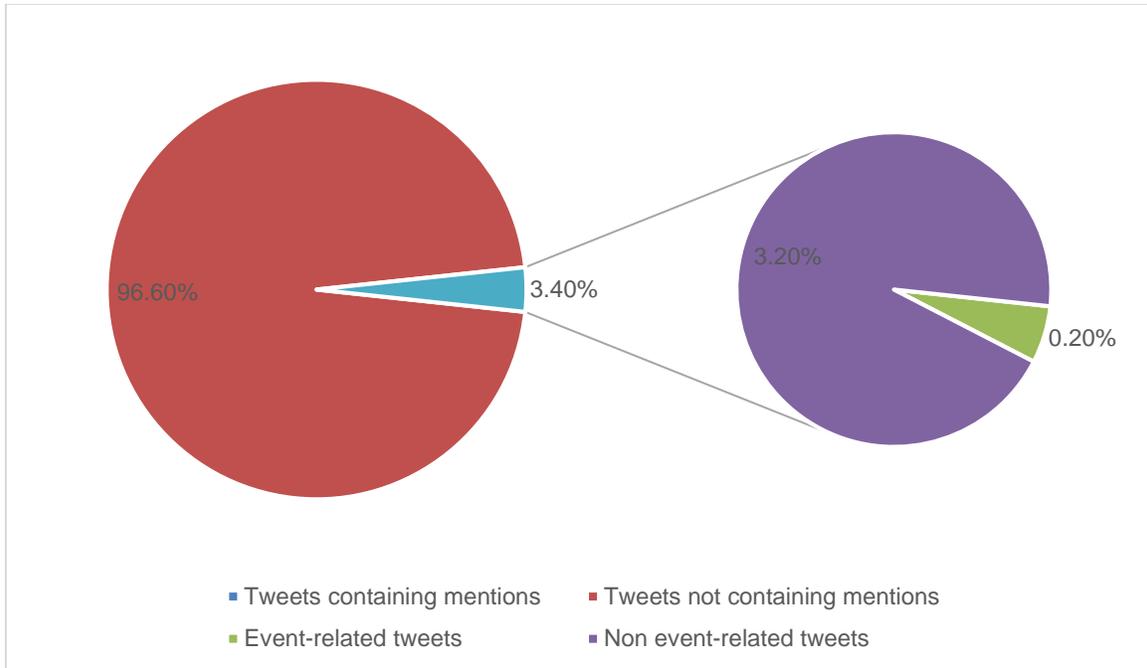

**Fig. 3** Studying the effect of tweets containing mentions on event detection

### 4.2 Results of studying location in tweet text and hashtags

In this study, two methods were combined to extract all the locations from the tweet text. The first method was utilizing the tweet text to extract location names by NER. The second method was investigating the tweet hashtags. The reason for utilizing both of these methods was that sometimes location names are mentioned only in hashtags or in both texts and hashtags. Fig. 4 shows the importance of using this combined method. As it can be seen in Fig. 4, 68.61% of location names have been mentioned only in tweet texts, 10.72 % only in hashtags, and 20.67% in both tweet texts and hashtags. Hence, for covering all the cases, a combination of these two methods was utilized.

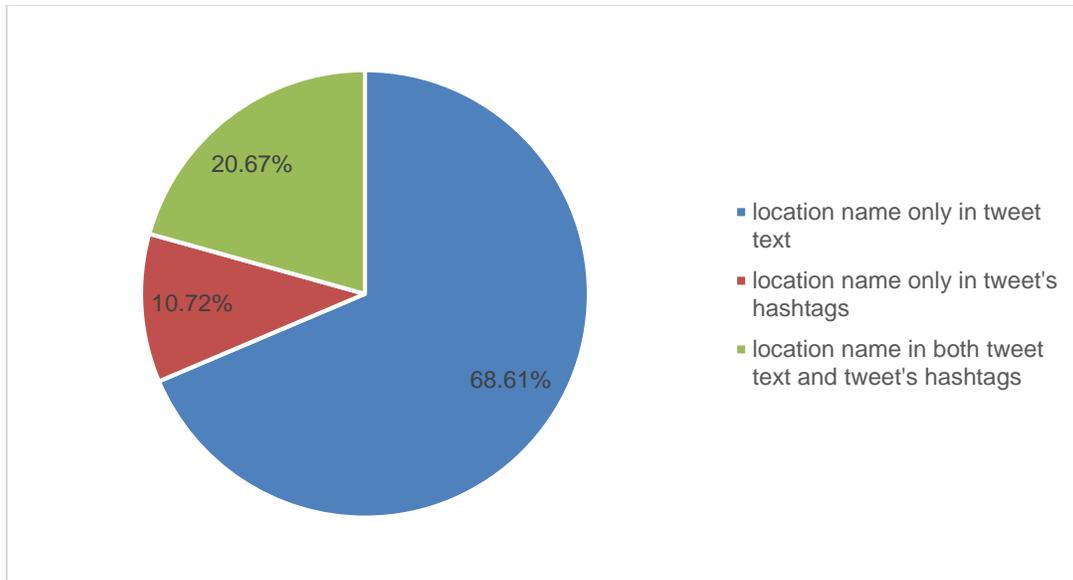

**Fig. 4** Results of studying location name in three cases

### 4.3 Results of event detection

Table 1 shows the results of event detection in the collected dataset. As it can be seen in Table 1, most of the events have been detected in both base and suggested methods, but a number of events have been detected only with the suggested method. These events were of less importance than the other events. This issue shows that the suggested method can detect all events, including hot events and less important ones.

**Table 1** Event detection results

| Date | Event keywords | Event description | Base method | Suggested method |
|---|---|---|---|---|
| 2019/7/31 | Remove-currency-zero | Removing 4 zeros from national currency | ✓ | ✓ |
| | Caspian sea | Protest against Iran's share in the Caspian sea | ✓ | ✓ |
| | Broadcast | Broadcasting day | ✗ | ✓ |
| 2019/8/1 | Sanction-Zarif-Foreign Minister | The US sanctions Iran's foreign minister | ✓ | ✓ |
| | #project2533 | 2533 civil projects of deprivation elimination in Sistan and Baluchestan | ✗ | ✓ |
| | #increase-capacity-threat-health #combat-congress-with-monopoly | Reactions to medical field capacity in universities | ✓ | ✓ |
| 2019/8/11 | Iraq-#Iraq_will_not_burn-#Iran_and_Iraq_can_not_be_separated | Reactions to demonstrations in Iraq | ✓ | ✓ |
| 2019/11/15 | petrol | Petrol rationing | ✓ | ✓ |
| 2020/1/3 | Assassination-General-Soleimani #harsh_revenge | The assassination of General Soleimani | ✓ | ✓ |
| 2020/1/8 | Missile-Base-Iran-#harsh_revenge | Iran missile attack to USA base in Iraq | ✓ | ✓ |
| | Ukrainian airplane | Ukrainian airplane crash | ✓ | ✓ |

Note: All the presented Event keywords in this table are translated from Persian to English.

### 4.4 Results of execution time

Based on Fig. 5, the execution speed of the event detection process is following a linear form. It is because of the equal number of extracted tweets on different days. On the other hand, the execution speed of the event detection process in the suggested method had some ups and downs, which caused by the difference in tweets processing time to investigate the location score. It is evident that if a tweet has more mentioned locations, it needs more time to be processed in the Gazetteer database. Chart peaks represent days in which most of its tweets mentioned many locations, and chart troughs represent days in which most of its tweets did not mention any location. According to Fig. 5, the average execution time of event detection has decreased in the suggested method than the base method. Overall, the execution time of event detection has improved 27% on average. One of its reasons was ignoring tweets containing mentions that were mostly noisy, and they were not event-related. Ignoring tweets containing mentions was a simple process that can help the execution speed of event detection significantly. Another reason for the improved execution time was utilizing the method which weighted the tweet's features for detecting event clusters.

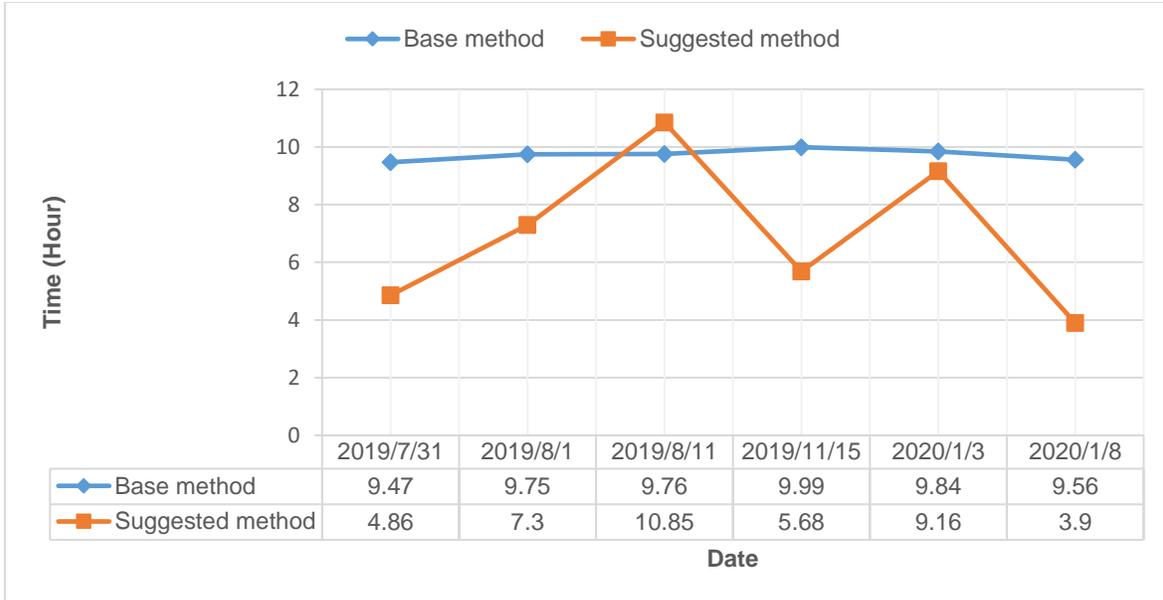

**Fig. 5** Comparing execution times

### 4.5 Results of validity (precision)

Precision is one of the criteria that has been considered for comparing the suggested method performance with the base method. Fig. 6 shows the results. Precision is defined as follow:

$$\text{Precision} = \frac{\text{TP}}{\text{TP}+\text{FP}} \qquad (4)$$

Fig. 6 shows the values of event detection precision in both suggested and base methods for each day separately. The results show that the suggested method increases the precision of event detection on average. The improvement happens because of ignoring tweets containing mentions and removing URLs from tweet texts.

The reason for the significant difference between the two methods on some dates, like 2019/7/31, was the existence of numerous tweets containing mentions in those dates. Hence, by ignoring these tweets in the suggested method, the effect of these kinds of noise can be prevented. As a result, the precision of event detection has improved 31% on average than the base method.

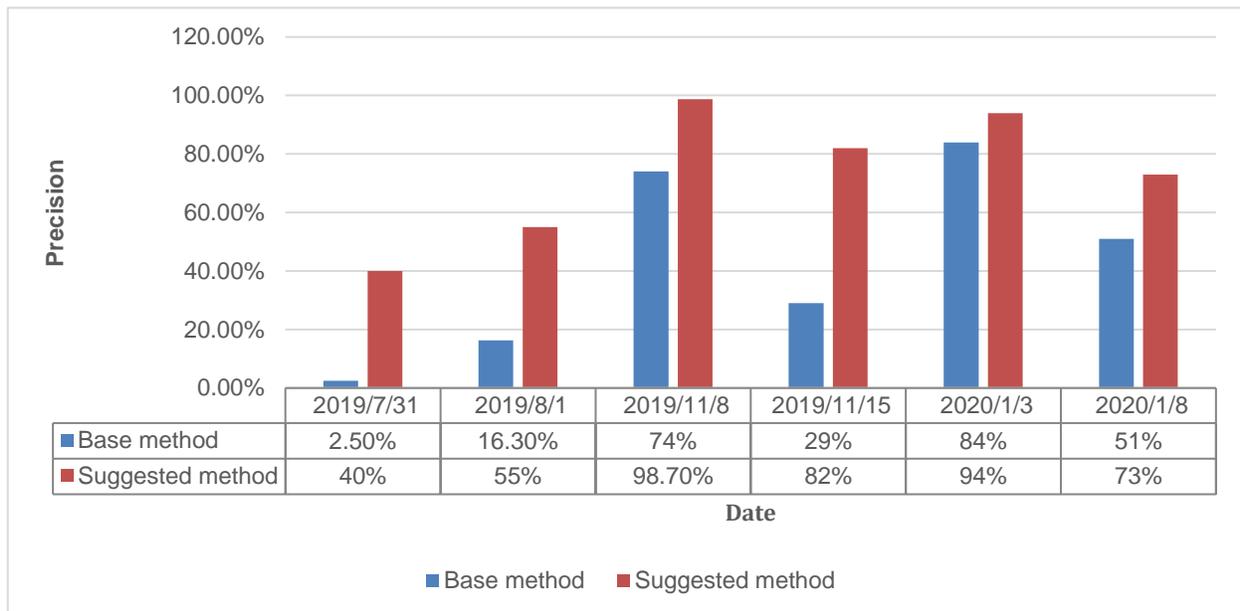

**Fig. 6** Comparing the precision of methods

## 5  CONCLUSION AND SUGGESTIONS

Different kinds of social networks that have been popular recently are rich sources of information. Information of these social networks can be used for detecting events so the authorities can use them to have more accurate and timely reactions in critical conditions. Twitter is one of the popular social networks which based on its entity and function, can be used for informing events. The main idea of this study was that based on tweet's features, all tweets should not be assigned the same weight. Hence, we tried to use the idea of weighting tweet's features for event detection. A tweet has different features, but in this study, we only studied three features, including the followers count, the retweets count, and the user location. Finally, by implementing the base and suggested methods, we concluded that the suggested method was able to detect all events, including hot and less important ones. The average execution time in the suggested method has improved 27% than the base method. Also, the average precision of event detection has improved 31% than the base method. Generally, it can be said that the suggested method had better performance than the base method.

Event detection is a domain with many potentials. For future researches, another social network like Telegram that is very popular among Iranians can be studied.

Another suggestion is using the combined information of several social networks. Also, the effect of other tweet's features in event detection can be studied.